# Coordinated Cyber-Attack Detection Model of Cyber-Physical Power System Based on the Operating State Data Link

**Lei Wang[1,2]\*, Pengcheng Xu[3], Zhaoyang Qu[1,2], Xiaoyong Bo[1,2] Yunchang Dong[1,2], Zhenming Zhang [1,2],Yang Li[1]**

[1]School of Electrical Engineering, Northeast Electric Power University, Jilin, China

[2]Jilin Engineering Technology Research Center of Intelligent Electric Power Big Data Processing, Jilin, China

[3]Siping Power Supply Company of State Grid Jilin Electric Power Company Limited, Siping, China

**\* Correspondence:**
Lei Wang
752953593@qq.com

**Abstract**

Existing coordinated cyber-attack detection methods have low detection accuracy and efficiency and poor generalization ability due to difficulties dealing with unbalanced attack data samples, high data dimensionality, and noisy data sets. This paper proposes a model for cyber and physical data fusion using a data link for detecting attacks on a Cyber-Physical Power System (CPPS). Two-step principal component analysis (PCA) is used for classifying the system's operating status. An adaptive synthetic sampling algorithm is used to reduce the imbalance in the categories' samples. The loss function is improved according to the feature intensity difference of the attack event, and an integrated classifier is established using a classification algorithm based on the cost-sensitive gradient boosting decision tree (CS-GBDT). The simulation results show that the proposed method provides higher accuracy, recall, and F-Score than comparable algorithms.

## 1    Introduction

In recent years, a new type of coordinated cyber-physical attack has caused blackouts of the power grid and disrupted power systems. The main reason is that the coordinated attack on the power grid by hackers was not detected in time, and effective measures to prevent major accidents could not be implemented at the optimum time (Haes et al., 2019; Lai et al., 2019). In the 2015 attack on the Ukrainian power grid, the attack point was not the power infrastructure, and the zero-day vulnerability was not used. Its attack cost is significantly lower than that of Stuxnet, Equation, and other attacks, but it is also more effective (Koopman et al., 2019; Zhang et al., 2016). Therefore, traditional security protection methods for power systems have their limitations, and it is urgent to research detection and defense methods for coordinated attacks on CPPS to identify attack types and intentions. It is crucial to establish a comprehensive active defense system to ensure the security of power systems (Chen et al., 2011; Dai et al., 2019; Wang et al., 2019b).

Many scholars have investigated the detection and identification of coordinated attacks on CPPS. The coupling relationship between the cyber side and the physical side has been considered in several studies (Drayer et al., 2019; Shen et al., 2019), which focuses on the fusion of the attack path on the information side and the attack object on the physical side. Xu et al. (2017) combined a priori and posterior bad data detection and proposed a new decomposition method to solve the state estimation data corruption in cyber-attacks. Kurt et al. (2018) and Basin et al. (2016) used a dynamic equation of the measured variables with a joint transformation to detect false data injection (FDI) attacks in real-time to improve the detection accuracy.

In summary, existing detection methods for cyber-attacks on CPPS have the following limitations: 1) the cyberspace and the physical space are closely coupled and interact with each other. An attack detection from the cyber side or the physical side alone is not sufficient (Lin et al., 2017; Nath et al., 2019). 2) Attack detection methods based on physical power grid data ignore the impact of cyber network attacks on the performance of smart grids. The effects of power grid failures and cyber-attacks on the physical side are similar, and it is difficult to distinguish them based on data characteristics (Huang et al., 2020; Liu et al., 2016). 3) A cyber-attack is characterized by unbalanced attack samples, high data dimensionality, and noise, and data with a long tail are common. Low detection accuracy of attacks and low real-time detection efficiency are typical (Osanaiye et al., 2018; Tian et al., 2019).

In this paper, the cyber-side alarm data and the physical-side measurement data are merged to establish a cyber-physical coupling state chain. A clustering method is designed to classify and distinguish different operating states of the CPPS. An oversampling algorithm is used to reduce the imbalance in the operating states' samples. Subsequently, a coordinated cyber-attack detection algorithm based on the improved gradient boosting decision tree (GBDT) is proposed. The algorithm optimizes the cost-sensitive (CS) loss function, minimizing the error associated with the small sample size of attack data and providing high accuracy of attack detection and a high recall rate and F1-score.

## 2 Detection Model for Coordinated Cyber-Attacks on CPPS

The framework of the coordinated cyber-attack detection model is shown in **Figure 1**. The model exploits the data characteristics in different states, such as normal operation, fault operation, and coordinated attack of the CPPS. First, the data link of the cyber-physical operation state is established according to the coupling relationship. A clustering algorithm is used to classify the state data link, and a feature set is obtained under different operating conditions. Then, the adaptive synthetic sampling algorithm (ADASYN) is used to balance the majority of the samples and the minority of the samples in different state data sets. Finally, new CS conditions are added using the GBDT's CS loss function to detect different coordinated cyber-attacks.

## 3 Establishment of the Data Link of the Cyber-physical Operating State

### 3.1 Data link of the operating state of the physical power grid

The physical grid measurement data reflects the real-time operating status of the grid under different working conditions. The measurement data of each section of the grid reflects the operating status at that moment. We do not consider the reasons for changes in the grid state (caused by cyber-attacks or general equipment failures); it can be described as a specific interval $\Delta t$ ($t_1 \sim t_n$). According to the acquisition sequence, all state data fragments $S_p(t_i)$ consisting of the physical grid operating data link $Q_p(\Delta t)$ are defined as follows:

$$\begin{cases} X_p(t_i) = \{x_1, \ x_2, \ldots x_h\} \\ S_p(t_i) = \{X_{p_1}, \ X_{p_2}, \ldots X_{p_m}\} \\ Q_p(\Delta t) = \{S_p(t_1), S_p(t_2) \ldots, S_p(t_n)\} \end{cases} \quad (1)$$

where $X_p(t_i)$ represents the $h$ measured attributes obtained from the physical-side device $X_p$ at time $t_i$, including the voltage, current, phase angle, active power, and reactive power; $S_p(t_i)$ represents all the measurement data collected by $m$ devices on the physical side at time $t_i$.

### 3.2 Data link of the operating state of the cyber network

The transmission delay and data packet loss rate typically reflect the performance status of a cyber network. When the control signal or status information is lost during the transmission of the data packet because it exceeds the allowable proportion, the control of the device has been lost due to a network attack (Davarikia et al., 2018; Wang et al., 2019a). Three indicators (delay rate, packet loss rate, and threat degree) are established to characterize the operating data link of the cyber network.

(1) The delay ratio (DR) is defined as follows:

$$R_{dr}(n) = \frac{\sum_{k=1}^{n} \left| \dfrac{P'_k}{P_k} - P_T \right|}{n} \times 100\% \quad (2)$$

where $n$ is the number of communication links transmitting data, $P'$ is the number of data packet losses for link $k$, $P_k$ is the number of data packets sent by link $k$, and $P_T$ is the threshold of the packet loss rate of link $k$.

(2) The packet loss ratio (PR) indicator is defined as follows:

$$R_{pr}(m) = \frac{\sum_{l=1}^{m}(T_l^{send} - T_l^{receive})}{m} \times 100\% \quad (3)$$

where $m$ is the number of devices that send information, $T_l^{send}$ is the sending time of data packet $l$, and $T_l^{receive}$ is the receiving time of data packet $l$.

(3) Threat degree $W_{th}(a_{i,j})$. Assuming that $n$ alarm events are generated within the sampling time window $\Delta t$, the address set of the information equipment is $\{IP_1, \ IP_2, \ \ldots IP_m\}$, and $a_{i,j}$ indicates that $IP_i$ contains $j$ alarm events. The intrusion detection system (IDS) deployed in the power cyber network indicates that the original threat degree is W. The threat degree is redefined as follows to determine the impact of alarm events on the attack risk of the entire system:

$$W_{th} = \frac{\sum_{i=1}^{n_i} \left| w_{ij} - \overline{w_i} \right|}{n_i} \times 100\% \tag{4}$$

where $w_{ij}$ is the threat degree of alarm events $a_{i,j}$, $w_i$ is the average value of the threat degrees of all alarm events in $IP_i$, and $n_i$ is the number of all alarm events in $IP_i$.

The three performance indicators of the operating status of the cyber network are used to establish the cyber system operating data link $Q_c(\Delta t)$ in the interval $\Delta t$ $(t_1 \sim t_n)$:

$$\begin{cases} Y_c(t_i) = \{R_{dr}, \quad R_{pr}, \quad W_{th}\} \\ S_c(t_i) = \{Y_{c1}, \quad Y_{c2}, \dots Y_{ck}\} \\ Q_c(\Delta t) = \{S_c(t_1), S_c(t_2), \dots, S_c(t_n)\} \end{cases} \tag{5}$$

where $Y_c(t_i)$ represents the $R_{dr}$, $R_{pr}$, and $W_{th}$ obtained from the cyber-side device $Y_c$ at time $t_i$; $S_c(t_i)$ represents the status data obtained from $k$ devices on the cyber side at time $t_i$.

### 3.3 Coupled mapping of the operating state of the cyber-physical system

We use topological mapping to couple and map the data links of the two heterogeneous networks to form a data link of the cyber-physical operating state. The grid can be divided into $m$ areas according to the physical grid connectivity, and each area has $n$ transmission lines. It is assumed that a line consists of $k$ electrical components $\{X_1, X_2, \dots, X_k\}$, each line is connected to $n$ communication devices $\{Y_1, Y_2, \dots, Y_n\}$, and each communication device has a unique IP address $\{IP_1, IP_2, \dots, IP_n\}$ in the cyber network. We sequentially connect each electrical component number, line number, and connected area in the data chain to create an index table linking the <connected area number Area, line number Line, electrical component ID number, and information component IP address>. The cyber network operating data link $Q_c$ and the physical power grid operating data link $Q_p$ in the interval are compared using the index table, and the data are stored in the corresponding index.

The cyber network clock with a collection period of T is used, and we set the sampling time window to $\varepsilon = [T - \alpha T, T]$, where $\alpha$ is the window size parameter. The larger the value, the longer the collection period is. In the sampling time window $\varepsilon$, many identical state events may occur in the cyber-physical coupling state chain. Therefore, these repetitive events are filtered and compressed to form the cyber-physical operating state data link, which is expressed as follows:

$$Q(\varepsilon) = \{x_1[Q_p(t_1), Q_c(t_1)], x_2[Q_p(t_2), Q_c(t_2)], \dots, x_n[Q_p(t_n), Q_c(t_n)]\} \tag{6}$$

## 4 Coordinated Cyber-Attack Detection Model of the CPPS

### 4.1 Operating state clustering based on two-step PCA

There are no labels for the different state categories in the original cyber-physical operating state data link $Q(\varepsilon)$. It is necessary to distinguish the different state categories using cluster analysis. In this paper, the two-step principal component analysis (PCA) clustering algorithm is proposed. The PCA algorithm is used to cluster, transform, and filter the correlated attributes to extract linear uncorrelated attributes (Jian et al., 2004). The two-step algorithm is used to cluster the attribute set; it reduces the computational complexity and provides high clustering accuracy (Northrup et al., 2004, Phelps et al., 2009, Dom et al., 2003). The algorithm steps are as follows:

Input: cyber-physical operating state data link $Q(\varepsilon)=\{x_1, x_2, \ldots, x_n\}$.

Output: D=$\{x_i, C_i\}$, where $C_i$ is the operating state of the clusters C=$\{C_1, C_2, \ldots, C_k\}$.

Step 1: Feature selection for clustering. The PCA algorithm is used to map $n$ attributes in the data link to $m$ dimension ($m<n$). The correlated attributes are filtered using an orthogonal transformation to obtain $m$-dimensional new features, A=$\{A_1, A_2, \ldots A_m\}$. The centralizing mean $x_i = x_i - \frac{1}{n}\sum_{i=1}^{n} x_i$ is used to derive the covariance matrix $XX^T$, whose eigenvalues and eigenvectors are obtained. The data link set $Q'(\varepsilon)$ is obtained after dimensionality reduction.

Step 2: Calculate the number of category clusters in the operating state. After the traversal process, the clustering feature (CF tree) growth in the balanced iterative reducing and clustering using hierarchies (BIRCH) algorithm is applied to the data link set $Q'(\varepsilon)$. The data points in the data set are evaluated one-by-one to collect all data points in the dense area while generating the CF tree. The log-likelihood distance $d(C_s, C_t) = \zeta_s + \zeta_t - \zeta_{<s,t>}$ between the two clusters is used to create many small sub-clusters. The Bayes information criterion (BIC) is used to calculate the number of possible division schemes for the state category.

Step 3: Determine the number of categories $C_J$ in the $Q'(\varepsilon)$. The agglomerative hierarchical clustering (AHC) method is used to merge the sub-clusters one-by-one, and the desired number of clusters is reached according to the $R(k)$ between the two clusters.

$$R(k) = \frac{d_{\min}(C_k)}{d_{\min}(C_{k+1})} \qquad (7)$$

where $C_k$ and $C_{k+1}$ is a partition scheme with $k$ or $k+1$ cluster numbers; $d_{min}(C_{k+1})$ and $d_{min}(C_k)$ is the distance between the two smallest clusters in the scheme.

Step 4: Label the sample data in each operating state cluster. The data points in each cluster are determined; the data points $x_i$ in the state data link set $Q'(\varepsilon)$ are regarded as single-point clusters according to the clustering results $C_J$. The logarithm similarity between $x_i$ and each cluster in $C_J$ is determined. Given the distance $d\{x_i\}, C_J\}$, $x_i$ is placed into the nearest cluster, and labels are generated for each operating state category C=$\{C_1, C_2, ..., C_k\}$.

## 4.2    Algorithm to reduce the imbalance of the operating state classes

A coordinated cyber-attack event of the CPPS has a small probability and high risk. In the data link $Q(\varepsilon)$, normal operation data account for the largest proportion, whereas the proportion of attack data is relatively small, resulting in unbalanced data. Therefore, the ADASYN algorithm is used to deal with the imbalance of the operating state classes (Wang et al., 2020; Qu et al., 2018). Balanced data distribution is obtained by adaptive synthetic oversampling. Different minority samples are given different weights to generate different numbers of samples. The algorithm process is as follows:

Input：$D=\{x_i, C_i\}$，where $x_i$ is the cyber-physical operating state data link $Q(\varepsilon)$，$C_i$ is the class label. $\alpha$ is the imbalance threshold, $C_k$ is a minority class, and $C_1$ is the majority class.

Output: Balanced data set $D'$.

Step 1: Calculate the class imbalance, where $\text{Imbalance} = \dfrac{\text{Lagre num}(C_l)}{\text{Small num}(C_k)}$. Calculate the number of samples to be synthesized based on the degree of imbalance
$G = (\text{Lagre num}(C_l) - \text{Small num}(C_k)) \times \beta,\ \beta \in [0,1]$.

Step 2: Calculate the proportion of the majority class in the $K$-nearest neighbors (KNNs).
$r_i = \Delta_i / K$，where $\Delta_i$ is the number of samples of the majority class in the KNN.

Step 3: Calculate the majority class surrounding each minority sample.

$$\hat{r_i} = \frac{r_i}{\sum\limits_{i=1}^{samll\ num(C_k)} r_i} \tag{8}$$

Step 4：Calculate the number of samples that need to be generated for each minority sample $C_k$.

$$g_i = \hat{r_i} \times G \tag{9}$$

Step 5: Select a minority sample among $k$ neighbors around each minority sample, and synthesize using Eq. (10). Repeat the synthesis $g_i$ times until the desired number of synthesized samples is obtained.

$$s_i = X_i + (X_{zi} - X_i) \times \eta \tag{10}$$

where $s_i$ is the composite sample, $X_i$ is the $i$-th sample in the minority sample, $X_i \in [0,1]$, $X_{zi}$ is a randomly selected minority sample among the KNNs of $X_i$.

Repeat the synthesis until the desired number of synthesized samples in Eq. (5) has been obtained.

### 4.3   Classification Algorithm of Coordinated Cyber−Attacks Based on CS−GBDT

The purpose of attack detection is to minimize the harm to the power grid caused by the attack. The harm caused by misinterpreting an attack as a normal event is far greater than that caused by misinterpreting a normal event as an attack (Huang et al., 2018). We propose using the CS function to improve the GBDT (Sakhnovich et al., 2011, Liao et al., 2016). The CS loss function replaces the standard cost loss function to prevent attack event misclassification. The improved CS loss function is defined as follows:

$$\text{Loss}(C, f(x)) = -\sum_{k=1}^{K} w_k C_k \log(p_k(x)) \tag{11}$$

where $K$ is the class of all attacks, $C_k$ is the sample of the $k$-th attack, and $p_k(x)$ is the probability of the $k$-th attack, $w_k$ is the cost-sensitive function, it can be divided into two costs, i.e., the missed detection cost $w_{(-,+)}(1 - p(x))$ , $p(x) \geq \dfrac{w_-}{w_+ + w_-}$ and the misdetection cost $w_{(+,-)}p(x)$, $p(x) < \dfrac{w_-}{w_+ + w_-}$.

Coordinated cyber-attack detection is a multi-classification task. A total of $K$ types of attacks are assumed. The sample $x$ in the cyber-physical operating state set is obtained, and the CS-GBDT algorithm is used to determine which class the $x$ sample belongs to. The specific steps of the algorithm are as follows：

Input: Balanced data set $D' = \{(x_1, C_1), (x_2, C_2), ..., (x_N, C_N)\}$ , loss function $Loss(C_k, f_k(x))$, and the number of classifiers $M_{..}$

Output: A strong learner for attack classification $F(x)$.

Step 1: Initialize $f_{k0}(x)=0$, the number of categories classified k=1,2,…$K$.

Step 2: Starting from $t$=1 to t=$M$, there are $M$ iterations in total, repeating the step 3 through 6, at last building $M$ classifiers.

Step 3: The one-hot code for each class $y_i$ is generated. We calculate the probability of sending the $k$-th attack sample $p_k(x)$.

$$p_k(x) = \frac{e^{f_k(x)}}{\sum_{l=1}^{K} e^{f_l(x)}}$$

(12)

Step 4: Start from $k$=1 to $k$=$K$, repeating the step 5 through 6, we generate $K$ different CART classification trees $f_1(x), f_2(x), …f_K(x)$.

Step5: Calculate the negative gradient of each class in the $m$ class, and obtain the negative gradient error of the $i$-th sample corresponding to category $k$ in the $t$-th iteration:

$$r_{ki} = C_{ki} - p_k(x) = -\left[ \frac{\partial Loss(C_k, f_k(x))}{\partial f_k(x)} \right], \;\; i=1, 2, \ldots, N$$

(13)

where $N$ is the number of sample data.

We use the estimated residual $\{(x_1, r_{k1}), … (x_N, r_{kN}))\}$ as an input to calculate the leaf node area of the $m$-th decision tree:

$$C_{mkj} = \frac{K\text{-}1}{K} \frac{\sum_{x_i \in R_{mkj}} r_{ki}}{\sum_{x_i \in R_{mkj}} |C_{ki}|(1 - |C_{ki}|)}$$

(14)

where $R_{mkj}$ is the leaf node region $R_{mj}$ of the m-th tree. $K$ is the number of categories.

Step 6: Update the classifier $f_{mk}(x)$.

$$f_{mk}(x) = f_{k,m-1}(x) + \sum_{j=1}^{J} C_{mkj} I \, , \, x \in R_{mkj} \tag{15}$$

where $J$ is the number of leaf nodes per tree.

Step 7: Build final classification tree with high accuracy used for attack detection.

$$F_{mk}(x) = \sum_{m=1}^{M} \sum_{j=1}^{J} C_{mkj} I \, , \, x \in R_{mkj} \tag{16}$$

## 5    Experimental analysis

### 5.1    Experimental environment and data

We simulate the different fault states of the physical power grid caused by cyber-attacks on the IEEE39-bus system in the RT-LAB and OPNET co-simulation environment. We collect the DR, PR, and threat information at different times on the cyber side. The voltage, current, impedance, and other data are collected on the physical side. The ten data sets are obtained. Each set contains 56 attributes, and the cumulative number of records is about 50000, including 5 types of operating states in the CPPS system, as follows:

1) Normal operating state (S1): there is no network attack on the cyber side, and the power grid on the physical side is operating normally. 2) Distributed denial of service (DDOS) attack state (S2): the data in the communication system are blocked by a DDOS attack, affecting the normal operation of the power system, measurement acquisition, and control commands. 3) Data injection attack state (S3): malicious data injection into physical power grid disguised as a normal fault, resulting in the operator mistakenly assuming a short-circuit fault. 4) Protection device parameter tampering attack state (S4): the attacker tampers with the distance parameter of the protection device, causing a failure of the protection device to disconnect the fault area. 5) Fault operation state (S5): the physical power grid has a single-phase, two-phase, or short-circuit fault.

### 5.2 Results of the operating state classification of the CPPS

The data set 1 with 4966 records is selected in the experiment. After implementing the two-step PCA clustering algorithm, the number of outliers is 89, and there are five operating states, as shown in **Figure 2(a)**. Cluster-3 (S2) has the largest number of records (1325). The clustering superiority is 0.93, and clustering importance is 0.85, accounting for 29.3% of the records. The smallest cluster is Cluster-4 (S5), with 59 records, accounting for 1.3% of all records. The clustering superiority is 0.97, and clustering importance is 0.93. Clustering superiority is a measure of cluster separation (-1~0.2 poor|0.2~0.5 medium|0.5~1 good), and clustering importance is a measure of cluster cohesion (0~0.2 poor|0.2~ 0.6 medium|0.6~1 good) (Nair et al., 1997).

According to the negative sequence current and zero sequence current amplitude of each cluster in the experiment, the curves of the three attack states and the fault state are obtained, as shown in **Figure 2(b)-(e)**. Cluster-2 is significantly different from the other four states, while Cluster-4 and Cluster-5 have high similarities. The reason is that Cluster-2 is an attack that causes network blocking and delay, which is significantly different from the other types of data tampering attacks.

Cluster-3 and Cluster-5 are physical power grid failures caused by information tampering attacks. These states are similar to the changes occurring in the Cluster-4 power grid normal fault.

The adjusted Rand index (ARI) is used to measure the accuracy of the clustering results; $\text{ARI} \in [-1, 1]$，The closer the value is to 1, the better the clustering performance is. The index is calculated as follows:

$$\text{ARI} = \frac{RI - E(RI)}{\max(RI) - E(RI)} \tag{17}$$

where $RI$ is the Rand coefficient, and $E(RI)$ is the expected value of each class.

Four typical clustering algorithms are selected for performance comparison, i.e., K-means, density-based spatial clustering of applications with noise (DBSCAN), clustering using representatives (CURE), and BIRCH. In the experiment, the sample size of the test data set is randomly selected and ranges from 5% to 100% of the data set. The ARI values of the different algorithms are shown in **Figure 2(f)**. As the proportion of the test data set increases, the ARI increases significantly. The accuracy of the proposed two-step PCA method is 97% for a sample size of 100%, demonstrating the excellent performance of this method. The K-means algorithm has the lowest ARI values.

### 5.3 Result of balancing the operation state classes

The number of samples in the operating state classes in 10 data sets before implementing the algorithm is shown in **Figure 3(a)**. The number of samples is imbalanced in the different operating states. The largest number of records (143766) occurs in the S3 state, and the fewest number (3080) is observed in the S5 state. The maximum class imbalance is 3.77. There are multiple minority and majority categories in the joint data set, showing multi-category imbalance.

The ADASYN algorithm is used to oversample the categories whose number is less than the threshold. We set the maximum imbalance threshold to 1.2. The results in the 10 data sets are shown in **Figure 3(b)**. The proportion of records in each dataset is close to 20%. The ADASYN algorithm uses local screening and sampling to reduce the influence of data imbalance on the false alarm rate of coordinated cyber-attack detection.

### 5.4 Performance verification of the coordinated cyber-attack detection in the CPPS

The balanced data set is divided into a training set (70% samples) and a test set (30% samples). The model loss parameters are set according to the improved CS loss function. There are 130 integrated base classifiers, and the depth of each independent tree (max_depth) is 7.

The receiver operating characteristics (ROC) curve obtained by classifying the test data set is shown in **Figure 4(a)**. The curves of the 5 categories are close to the (0,1) position, and the average area under the ROC curve (AUC) is 0.982. This result shows that the attack detection model has a low false alarm rate and high accuracy.

The precision-recall (PR) curve obtained by classifying the test data set is shown in **Figure 4(b)**. The PR curves are all close to the (1,1) position, indicating that the attack detection model has high recall and accuracy, even when the ratio of positive and negative samples is large. Therefore, the proposed attack detection model has a high classification accuracy for unbalanced data.

The confusion matrix of the attack detection results is shown in **Figure 4(c)**. The detection accuracy for the DDOS blocking attack (S2) is 98%, that of the data injection attack (S3) is 96%, that of the protection device parameter tampering attack (S4) is 97%, that of the normal operation (S1) is 99%, and that of the fault operation (S5) is 98%. These results demonstrate that the proposed coordinated cyber-attack detection model accurately detects coordinated attack events on the network and distinguishes attack states from the fault operation state, with a maximum false-positive rate of only 4%.

Finally, the proposed model is compared with typical classification algorithms, including the KNN, Xgboost, Random Forest, Adaboost, and support vector machine (SVM). The overall accuracy, average recall, average precision, and average F1-score of the algorithms are shown in **Figure 4(d)**. The recall and precision of the CS-GBDT algorithm are higher than 97%. The algorithm performance is stable, and it provides better performance for detecting various attack events than comparable algorithms.

## 6 Conclusion

In this paper, a cyber-physical operating state data link was established using data fusion mapping. The two-step PCA clustering algorithm is proposed for accurate labeling of the different operating states of the network. A coordinated cyber-attack classifier based on the CS-GBDT was established that considers the imbalance of the attack status categories and the cost sensitivity of the attack event. The algorithm can detect attacks on the CPPS and distinguish different attack types. The proposed model has a low false-alarm rate and high accuracy for attack detection. It is suitable for the detection of coordinated cyber-attack events with unbalanced attack sample data and high data dimensionality.

## 7 Data Availability Statement

The original contributions presented in the study are included in the article/supplementary material, further inquiries can be directed to the corresponding author/s.

## 8 Author Contributions

LW: designed the model framework of the paper and experimental verification. PX: contributed to the construction method of cyber physical operation state data link. ZQ: contributed to design the TwoStep PCA algorithm for operating state clustering. XB: completed the simulation experiment of attack classification detection. YD: performed the data collection and researched balance processing algorithm of operating state classes; ZZ: studied the classification algorithm of cyber cooperative attack based on CS_GBDT; YL: built a simulation environment and improved the grammar and sentence structure of the full paper.

## 9 Funding

This paper was supported in part by science and technology innovation development plan project of Jilin (20200401097GX).

## 10 Conflict of Interest

Author PX was employed by Siping Power Supply Company of State Grid Jilin Electric Power Company Limited, China.

The remaining authors declare that the research was conducted in the absence of any commercial or financial relationships that could be construed as a potential conflict of interest.

**Figures And Tables**

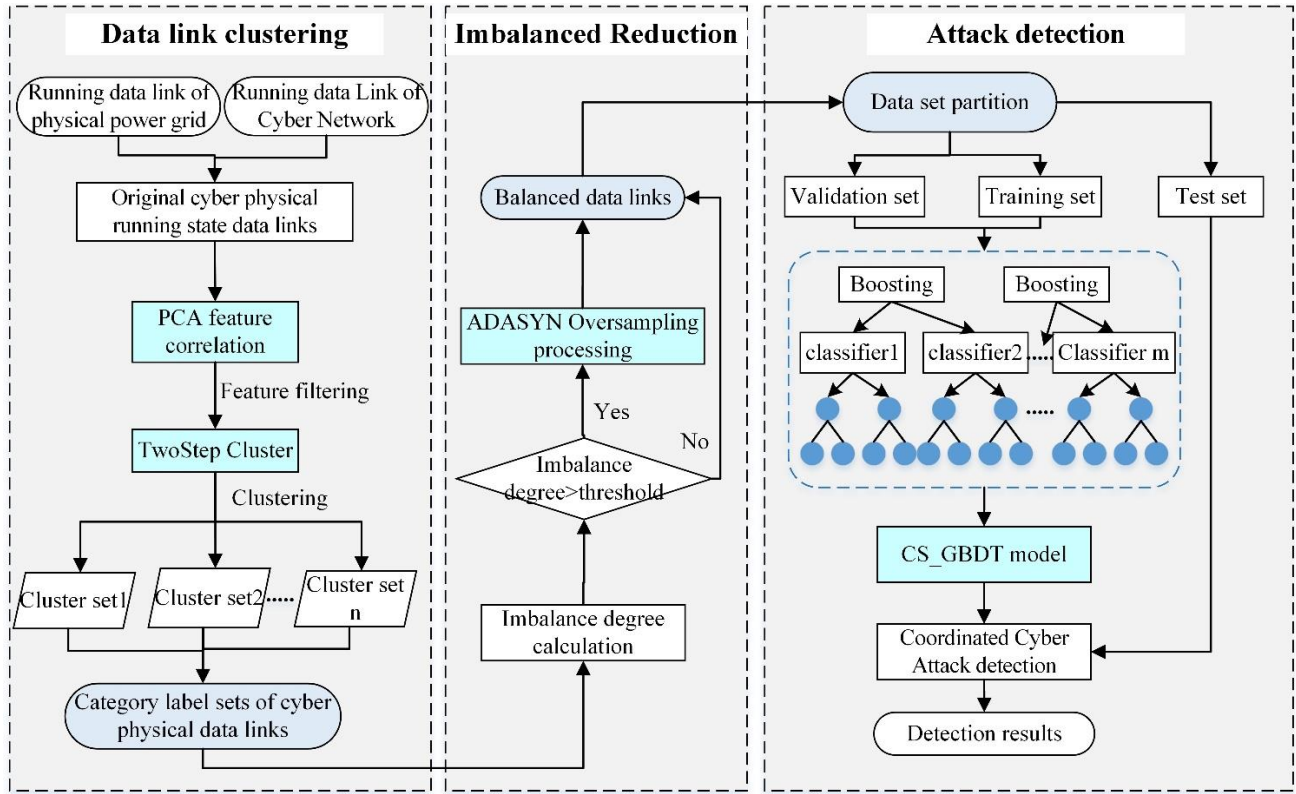

**Figure 1** Framework of the coordinated cyber-attack detection model for the CPPS

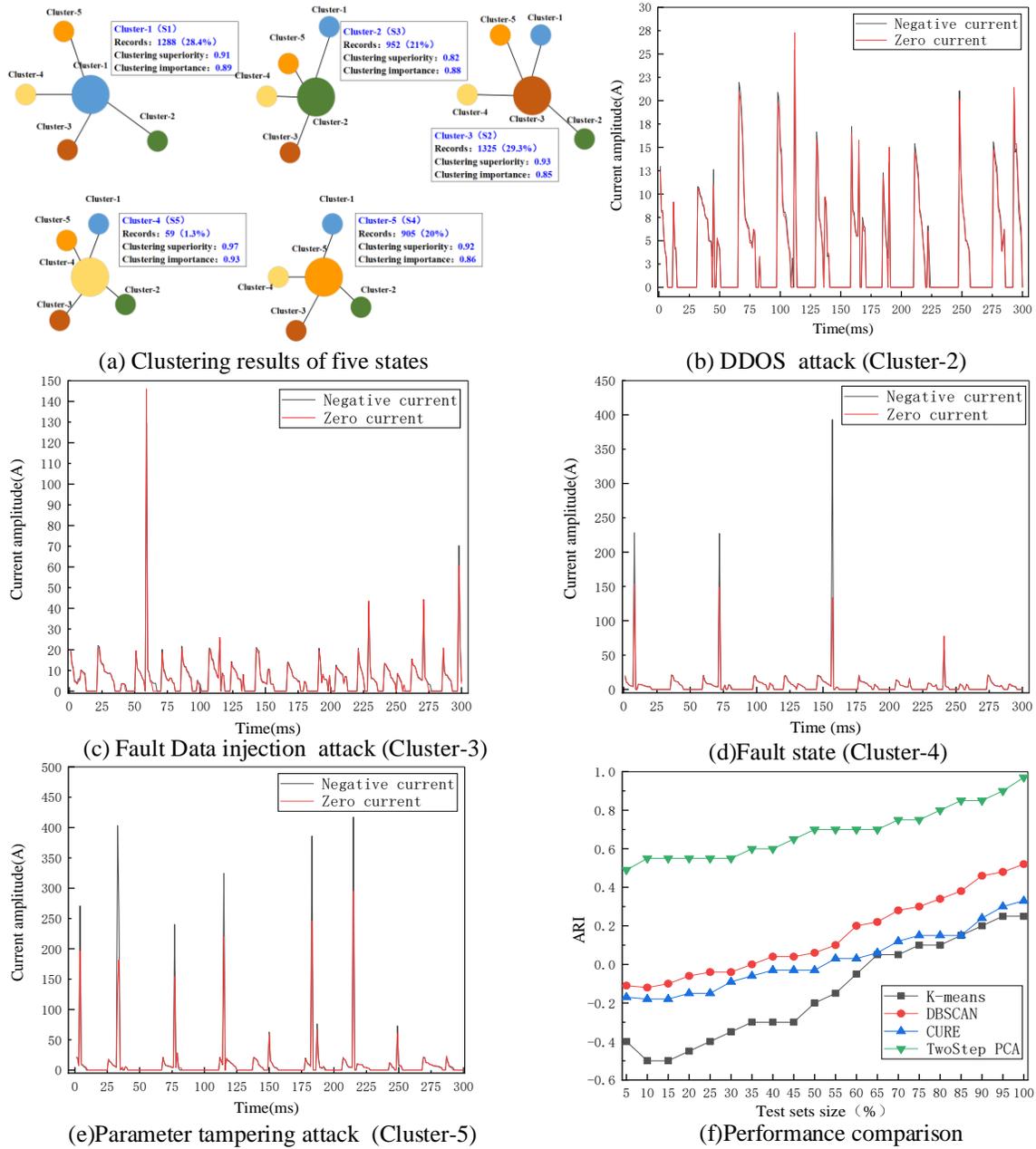

(a) Clustering results of five states

(b) DDOS attack (Cluster-2)

(c) Fault Data injection attack (Cluster-3)

(d)Fault state (Cluster-4)

(e)Parameter tampering attack (Cluster-5)

(f)Performance comparison

**Figure 2** Results of clustering the operating states and performance comparison of different clustering algorithms

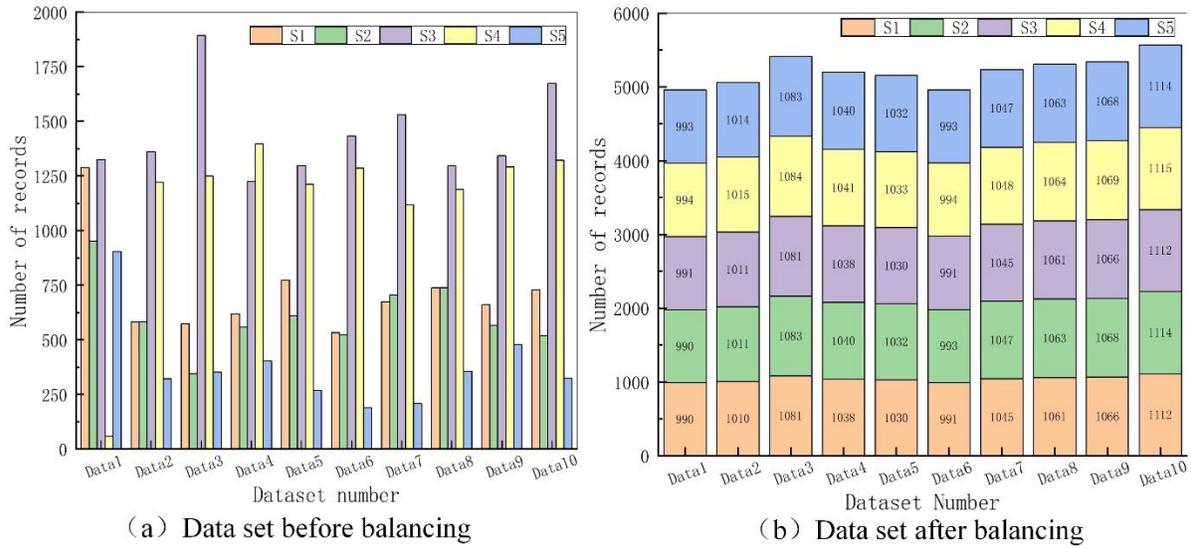

（a）Data set before balancing  　　　　　　（b）Data set after balancing

**Figure 3** Reducing the imbalance in the operating state categories

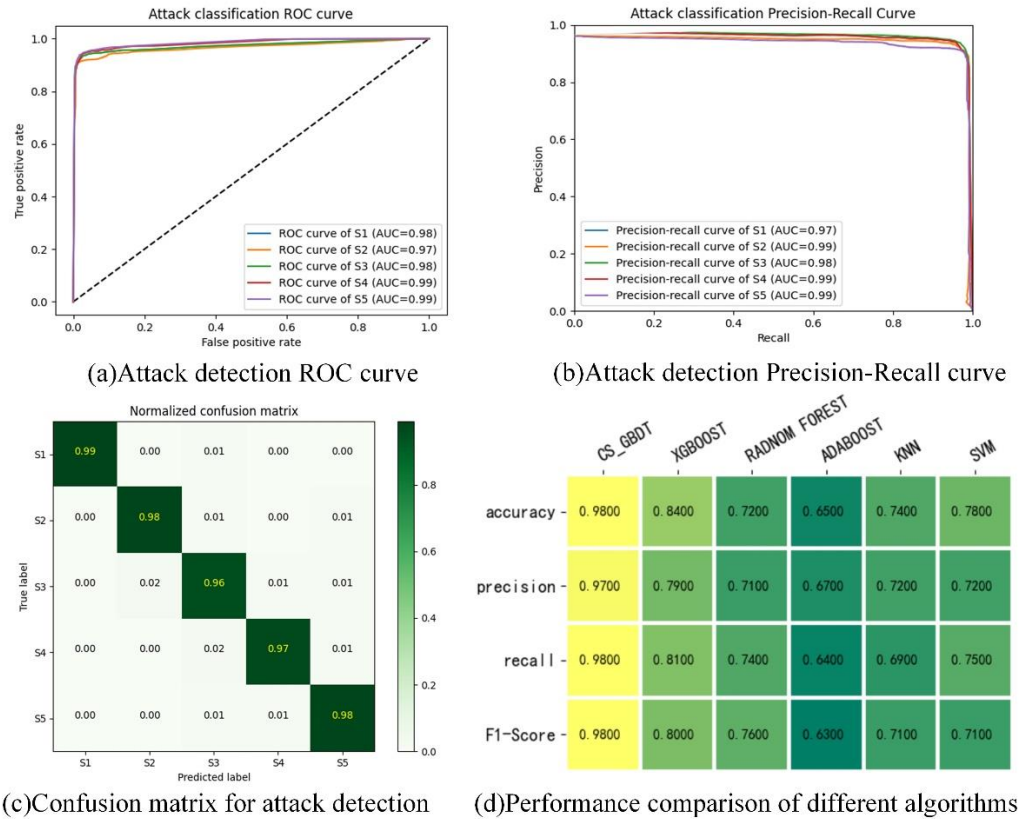

(a)Attack detection ROC curve  　　　　　(b)Attack detection Precision-Recall curve

(c)Confusion matrix for attack detection  　　(d)Performance comparison of different algorithms

**Figure 4** Attack detection ROC curve (a), Precision-Recall curve (b), Confusion matrix (c), and Performance comparison (d)